\documentclass[aps,showpacs,prl,toolkits,preprint]{revtex4}
\usepackage{epsfig,amssymb,amsfonts,graphicx}

\makeatletter
\renewcommand{\@makefntext}[1]{\parindent=1em\noindent\hbox to 1.8em
{\hss$^{\@thefnmark}$}#1}
\renewcommand{\@footnotemark}{\hbox{\mathsurround=0pt$^{\@thefnmark}$}}

\makeatother

\begin{document}
\title{Effective restoration of chiral symmetry in excited mesons.}
\author{R. F. Wagenbrunn    and L. Ya. Glozman}
\affiliation{Institute for
 Physics, Theoretical Physics branch, University of Graz, Universit\"atsplatz 5,
A-8010 Graz, Austria}

\newcommand{\be}{\begin{equation}}
\newcommand{\bea}{\begin{eqnarray}}
\newcommand{\ee}{\end{equation}}
\newcommand{\eea}{\end{eqnarray}}
\newcommand{\ds}{\displaystyle}
\newcommand{\low}[1]{\raisebox{-1mm}{$#1$}}
\newcommand{\loww}[1]{\raisebox{-1.5mm}{$#1$}}
\newcommand{\lmn}{\mathop{\sim}\limits_{n\gg 1}}
\newcommand{\vpint}{\int\makebox[0mm][r]{\bf --\hspace*{0.13cm}}}
\newcommand{\too}{\mathop{\to}\limits_{N_C\to\infty}}
\newcommand{\vp}{\varphi}
\newcommand{\vx}{{\vec x}}
\newcommand{\vy}{{\vec y}}
\newcommand{\vz}{{\vec z}}
\newcommand{\vk}{{\vec k}}
\newcommand{\vq}{{\vec q}}
\newcommand{\vpp}{{\vec p}}
\newcommand{\vn}{{\vec n}}
\newcommand{\vg}{{\vec \gamma}}

\begin{abstract}
A fast restoration of chiral symmetry in excited mesons is demonstrated.
A minimal "realistic" chirally symmetric confining model is used, where
the only interaction between quarks is the linear instantaneous Lorentz-vector 
confining potential. 
Chiral symmetry breaking is generated via the nonperturbative
resummation of valence quarks self-energy loops and the meson bound states
are obtained from the Bethe-Salpeter equation. The excited mesons fall
into approximate chiral multiplets and lie on
the approximately linear  radial and angular Regge trajectories, though
a significant deviation from the linearity of the angular trajectory
is observed.
\end{abstract}

\maketitle

There are certain phenomenological evidences that in highly excited hadrons,
both in baryons \cite{G1,CG,G2} and mesons \cite{G3,G4} chiral and $U(1)_A$
symmetries are approximately restored, for a short overview see \cite{G5}. 
This "effective"
restoration of chiral and $U(1)_A$ symmetries should not be confused with the
chiral symmetry restoration at high temperatures and/or densities.  
 What actually happens is that the excited hadrons gradually
decouple from the quark condensates. Fundamentally it happens because in the
high-lying hadrons the semiclassical regime is manifest and semiclassically
quantum fluctuations of the quark fields 
are suppressed relative to the classical 
contributions which preserve both chiral and $U(1)_A$ symmetries \cite{G5,G6}.
The microscopical reason is that in high-lying hadrons a typical momentum
of valence quarks is large and hence they decouple from the quark
condensate and consequently their Lorentz-scalar dynamical
mass asymptotically vanishes \cite{G1,G2,G6,G7,KNR}.
Restoration of chiral symmetry requires a decoupling of states
from the Goldstone bosons \cite{G2,CG2,JPS,G7} which is indeed observed
phenomenologically since the coupling constant for
$h^* \rightarrow h +\pi$
decreases fast higher in the spectrum.

At the moment there are two main paths to understand this phenomenon. In the
first one one tries to connect the high-lying states to the short-range
part of the two-point correlation function where the Operator Product Expansion
is valid \cite{CG,OPE}. However, the OPE is an asymptotic expansion.
Then, while the correct spectrum of QCD must be consistent with the
OPE, there is an infinite amount of incorrect spectra that can also
be in agreement with the OPE. Hence the results within the present
approach crucially depend on additional assumptions \cite{SHIFMAN,GOL}.

 In the 
second approach the 
authors try to understand this phenomenon within the microscopical 
models \cite{KNR,G6,G7,CG2}.
There are also interesting attempts to formulate the problem on the 
lattice \cite{DEGRAND,COHEN},
though  extraction of the high-lying states on the lattice is a 
task of  future.

In the absence of the controllable analytic solutions to QCD an insight 
into phenomenon
can be achieved only through  models. Clearly the model must be
field-theoretical (in order to be able to exhibit the spontaneous
breaking of chiral symmetry),
 chirally
symmetric and contain  confinement. 
In principle
any possible gluonic interaction can contribute to chiral symmetry breaking 
and it is not known which specific interaction is the most important 
one in this respect. 
However,
at the first stage it is reasonable to restrict oneselves to the simplest 
possible
model that contains all three key elements. Such a model is known, it is a
generalized Nambu and Jona-Lasinio model (GNJL) with the instantaneous 
Lorentz-vector
confining kernel \cite{Finger:1981gm,Orsay,BR}. This model 
is similar in spirit to
 the large $N_c$ `t Hooft model (QCD in 1+1 dimensions) \cite{HOOFT}.
In both models the only interaction between quarks is the instantaneous
infinitely raising Lorentz-vector linear potential. Then chiral symmetry breaking
is described by the standard   summation of the valence quarks
self-interaction loops in the rainbow approximation (the Schwinger-Dyson or gap
equations), while mesons are obtained from the Bethe-Salpeter equation for the
quark-antiquark bound states, see Fig. 1.
\begin{figure}
\includegraphics[width=0.6\hsize]{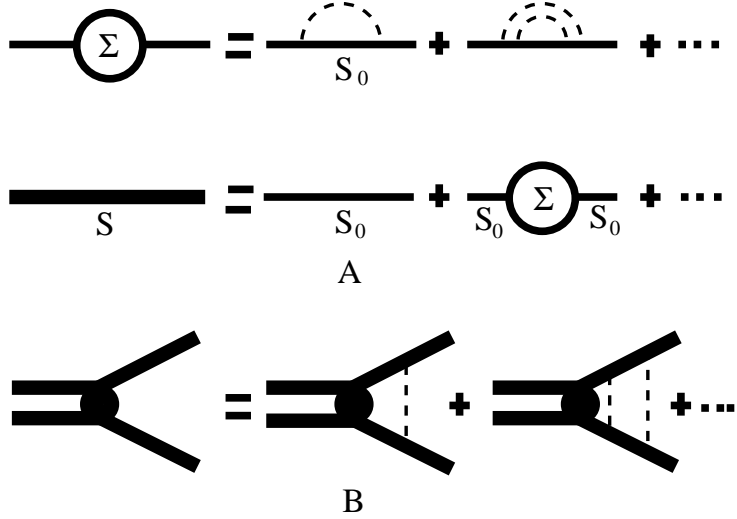}
\caption{ Graphical representation of the Schwinger-Dyson (A) and
Bethe-Salpeter equations  (B) in the ladder approximation.}
\end{figure}

Conceptually the underlying physics is very clear in the `t Hooft model 
in the sense
that once the proper gauge is chosen, the linear Lorentz-vector confining 
potential
appears automatically as the Coulomb interaction in 1+1 dimensions. 
In 3+1 dimensions,
once the Coulomb gauge is used for the gluonic field \cite{CL}, an 
almost linearly
raising confinement potential has been obtained \cite{Zwanziger:2003de,SS,R}.

An obvious advantage of the GNJL model is that it can be applied in 3+1
dimensions to systems of arbitrary spin.
In 1+1 dimensions there is no spin, the rotational motion of quarks 
is impossible,
and the states are characterized by the only quantum number, which is 
the radial
quantum number. Then it is known that the spectrum represents an alternating
sequence of positive and negative parity states and chiral multiplets never
emerge. This happens because in 1+1 dimensions the valence quarks
can perform only an oscillatory motion. In 3+1 dimension, 
on the contrary, the quarks can rotate and hence
can always be ultrarelativistic
and chiral multiplets should emerge naturally \cite{G2}.

Restoration of chiral symmetry in excited heavy-light mesons has been
studied with the quadratic confining potential \cite{KNR} and it was
also mentioned in a model with
the instantaneous potential of a more complicated form \cite{SW}.
Here we report our  results for excited light-light mesons with
the linear  potential. To our best knowledge this is the first
explicit demonstration of the restoration of the chiral symmetry
within the solvable "realistic" field-theoretical model. The results
for the lowest mesons within the similar model have been previously
reported in ref. \cite{COT}.

The GNJL model is described by the Hamiltonian \cite{Orsay}
\begin{eqnarray} 
\hat{H} & = & \int d^3x\bar{\psi}(\vec{x},t)\left(-i\vec{\gamma}\cdot
\vec{\bigtriangledown}+m\right)\psi(\vec{x},t) \nonumber \\
 &+& \frac12\int d^3
xd^3y\;J^a_\mu(\vec{x},t)K^{ab}_{\mu\nu}(\vec{x}-\vec{y})J^b_\nu(\vec{y},t),
\label{H} 
\end{eqnarray} 
%
with the quark current--current 
($J_{\mu}^a(\vec{x},t)=\bar{\psi}(\vec{x},t)\gamma_\mu\frac{\lambda^a}{2}
\psi(\vec{x},t)$) interaction parametrised by an instantaneous 
confining kernel $K^{ab}_{\mu\nu}(\vec{x}-\vec{y})$ of a generic
form. In this paper, we use the linear confining potential,
\begin{equation} 
K^{ab}_{\mu\nu}(\vec{x}-\vec{y})=g_{\mu 0}g_{\nu 0}
\delta^{ab} V_0 (|\vec{x}-\vec{y}|),
\label{KK}
\end{equation}
%
and absorb the color Casimir factor into string tension,
$\frac{\lambda^a \lambda^a}{4}V_0(r) = \sigma r$.

The Schwinger-Dyson equation for the self-energy operator 
$\Sigma(\vec p) =[A_p-m]+(\vec{\gamma}\hat{\vec{p}})[B_p-p]$
is
\begin{equation}
i\Sigma(\vec{p})=\int\frac{d^4k}{(2\pi)^4}V(\vec{p}-\vec{k})
\gamma_0\frac{1}{S_0^{-1}(k_0,\vk)-\Sigma(\vk)}\gamma_0, 
\label{Sigma03} 
\end{equation}
%
where
\begin{equation}
V(\vpp)=-\int d^3 x e^{i{\vpp\vx}} \sigma
|\vx|=\frac{8\pi\sigma}{p^4},
\label{FV} 
\end{equation}
%
so that the dressed Dirac operator becomes
\begin{equation}
D(p_0,\vec{p})= i S^{-1}(p_0,\vec{p}) =
\gamma_0p_0-(\vec{\gamma}\hat{\vec{p}})B_p-A_p,
\label{SAB}
\end{equation}
%
where, due to the instantaneous nature of the interaction the time-component
of the Dirac operator is not dressed. The Lorentz-scalar dynamical mass
$A_p$ as well as the Lorentz-vector spatial part $B_p$
contain both the classical and quantum contributions, the latter
coming from loops \cite{G6}:
\begin{eqnarray}
A_p & = & m+\frac{1}{2}\int\frac{d^3k}{(2\pi)^3}V
(\vec{p}-\vec{k})\sin\vp_k,\quad  \\
B_p & = & p+\frac{1}{2}\int \frac{d^3k}{(2\pi)^3}\;(\hat{\vec{p}}
\hat{\vec{k}})V(\vec{p}-\vec{k})\cos\vp_k, 
\label{AB} 
\end{eqnarray} 
%
where $\tan\vp_p=\frac{A_p}{B_p}$. 

Solution of the Schwinger-Dyson
equation (\ref{Sigma03})  with the linear potential is well-known, 
see e.g. \cite{BN}, and the mass-gap equation has a
nontrivial solution which breaks chiral symmetry, by generating
a nontrivial dynamical mass function $A_p$. This dynamical mass is
a very fast decreasing function at larger momenta. Then the
quark condensate is given as
\begin{equation}
\langle\bar{q}q\rangle=-\frac{N_C}{\pi^2}\int^{\infty}_0 dp\;p^2\sin\vp_p.
\label{Sigma1}
\end{equation} 

The homogeneous Bethe--Salpeter equation for the quark-antiquark bound
state with mass $M$ in the rest frame, i.e. with the four momentum
$P^\mu=(M,\vec{P}=0)$, is
\begin{eqnarray}
\chi(\vpp,M)&= &- i\int\frac{d^4k}{(2\pi)^4}V(\vpp-\vk)\;
\gamma_0 S(k_0+M/2,\vk) \nonumber \\
& \times & \chi(\vk,M)S(k_0-M/2,\vk)\gamma_0,
\label{GenericSal}
\end{eqnarray}
%
where $\chi({\vec p},M)$ is the mesonic Salpeter amplitude in the rest frame.
Eq.~(\ref{GenericSal}) is written in the
ladder approximation for the vertex which is consistent with the 
rainbow approximation for the quark mass operator and which is well
justified in the large-$N_C$ limit.

The Salpeter amplitude can be decomposed into two components for
mesons with $J^{PC}=(2n)^{-+}$, $(2n+1)^{+-}$, $(2n+1)^{++}$, $(2n+2)^{--}$,
or $0^{++}$ and into four components for mesons with
$J^{PC}=(2n+1)^{--}$ or $(2n+2)^{++}$, respectively. Here $J$ is the spin,
$P$ the parity and $C$ the charge conjugation parity of the meson and
$n\in {\mathbb N}_0$. In that way the Bethe--Salpeter equation becomes a system of
coupled integral equations for the components which we solve by expanding
them into a finite number of properly chosen basis functions. This leads to
a matrix eigenvalue problem which can be solved by standard linear algebra
methods. We vary the meson mass until one of the eigenstates is equal to one.

In the gap  as well as in the Bethe-Salpeter equations the  infrared
divergences are removed by introducing a finite "mass" into a
confining potential, which is a standard trick. Then  in the infrared limit 
("mass" goes to 0) the quark propagator consists of a finite and
diverged parts, while the mesons masses are finite.  Recently it was
demonstrated that also the masses of quark-quark subsystems in the 
color-antitriplet state go to infinity in this limit and  hence are removed
from the physical spectrum \cite{AL}. The results presented in the
following were obtained by calculating the infrared limit numerically,
i.e. the quoted meson masses were extrapolated to the infrared limit from
a few points with a very small but finite mass of the infrared regulator.
It turned out that in this region of the small mass of the infrared
regulator the squares of the meson masses depend almost linearly
on the mass of the infrared regulator making the extrapolation reliable.
The presented results are accurate within the quoted digits at least 
for states with small $J$ and for states with higher $J$ but small $n$. At
larger $J$ for larger $n$ numerical errors accumulate in the second digit
after comma.

By definition
an effective chiral symmetry restoration means that (i) the states
fall into approximate multiplets of $SU(2)_L \times SU(2)_R$ 
and the splittings within the multiplets ( $\Delta M = M_+ -M_-$) 
vanish at $n \rightarrow \infty$ and/or $ J \rightarrow \infty$ ;
(ii) the splitting within the multiplet is much smaller
than between the two subsequent multiplets  \cite{G3,G4,G5}.

The condition (i) is very restrictive, because the structure of
the chiral multiplets for the $J=0$ and $J>0$ mesons 
is very different \cite{G3,G4}.
For the $J>0$ mesons chiral symmetry requires a {\it doubling} of states
with some quantum numbers in contrast to the $J=0$ states. 
Given the complete set of standard quantum numbers $I, J^{PC}$,
the  multiplets of  $SU(2)_L \times SU(2)_R$  are

\begin{center}
{\bf J~=~0}

\begin{eqnarray}
(1/2,1/2)_a  &  :  &   1,0^{-+} \longleftrightarrow 0,0^{++}  \nonumber \\
(1/2,1/2)_b  &  : &   1,0^{++} \longleftrightarrow 0,0^{-+} ,
\end{eqnarray}
\bigskip
{\bf J~=~2k,~~~k=1,2,...}
\begin{eqnarray}
 (0,0)  & :  &   0,J^{--} \longleftrightarrow 0,J^{++}  \nonumber \\
 (1/2,1/2)_a  & : &   1,J^{-+} \longleftrightarrow 0,J^{++}  \nonumber \\
 (1/2,1/2)_b  & : &   1,J^{++} \longleftrightarrow 0,J^{-+}  \nonumber \\
 (0,1) \oplus (1,0)  & :  &   1,J^{++} \longleftrightarrow 1,J^{--} 
\end{eqnarray}

\bigskip
{\bf J~=~2k-1,~~~k=1,2,...}

\begin{eqnarray}
 (0,0)  & :  &   0,J^{++} \longleftrightarrow 0,J^{--}  \nonumber \\
 (1/2,1/2)_a  & : &   1,J^{+-} \longleftrightarrow 0,J^{--}  \nonumber \\
 (1/2,1/2)_b  & : &   1,J^{--} \longleftrightarrow 0,J^{+-}  \nonumber \\
 (0,1) \oplus (1,0)  & :  &   1,J^{--} \longleftrightarrow 1,J^{++} 
\end{eqnarray}
\end{center}

 Note that 
within the present
model the axial anomaly is absent. Even
so there are no exact $U(1)_A$ multiplets, because this
symmetry is broken not only by the anomaly, but also by the
chiral condensate of the vacuum. Then  the mechanism of the
$U(1)_A$ symmetry breaking and restoration 
is exactly the same as of $SU(2)_L \times SU(2)_R$. 
Hence the effective restoration of $SU(2)_L \times SU(2)_R$ 
would automatically imply restoration of $U(1)_A$ and of
$U(2)_L \times U(2)_R$ and vice versa.
An effective restoration of the $U(1)_A$ symmetry would mean 
an approximate degeneracy of
the opposite spatial parity states with the same
isospin from the distinct $(1/2,1/2)_a $ and $(1/2,1/2)_b$
 multiplets of $SU(2)_L \times SU(2)_R$. 
 
 Note that within the
 present model there are no vacuum fermion loops. Then since
 the interaction between quarks is flavor-blind the states
 with the same $J^{PC}$ but different isospins from the distinct
 multiplets $(1/2,1/2)_a $ and $(1/2,1/2)_b$ as well as the states
 with the same $J^{PC}$ but different isospins from $(0,0)$ and
 $(0,1) \oplus (1,0)$ representations are exactly degenerate. Hence it
 is enough to show a complete set of the isovector (or isoscalar)
 states.

In Table 1 we present our results for the spectrum for the two-flavor
($u$ and $d$) mesons in the chiral limit. Clearly the model should
not be taken seriously for the low-lying states  
where other gluonic interactions as well as 
the $1/N_c$ corrections should be important. The purpose of the study
is, however, to demonstrate that a solvable field-theoretical model 
in 3+1 dimensions does exhibit the effective
restoration of the chiral symmetry at large radial excitations $n$ and large
spins. The excited mesons fall into approximate chiral and $U(1)_A$
multiplets and all conditions of the effective symmetry restorations
are satisfied.
We observe  a very fast restoration of
both $SU(2)_L \times SU(2)_R$ and $U(1)_A$ symmetries with increasing
$J$ and essentially more slow restoration with increasing of $n$.
\begin{table}
\caption{Masses of isovector mesons in units of $\sqrt{\sigma}$.}
\begin{tabular}{c@{\hspace*{1.5em}}c@{\hspace*{1.5em}}rrrrrrr}
\hline\hline
\multicolumn{1}{c}{chiral}&\raisebox{-1.5ex}{$J^{PC}$}&
\multicolumn{7}{c}{radial excitation $n$}\\[-1.ex]
\multicolumn{1}{c}{multiplet}&&0\hspace*{0.7em}&1\hspace*{0.7em}&2\hspace*{0.7em}&3\hspace*{0.7em}&
4\hspace*{0.7em}&5\hspace*{0.7em}&6\hspace*{0.7em}\\
\hline
$(1/2,1/2)_a$&$0^{-+}$&0.00&2.93&4.35&5.49&6.46&7.31&8.09\\
$(1/2,1/2)_b$&$0^{++}$&1.49&3.38&4.72&5.80&6.74&7.57&8.33\\
\hline
$(1/2,1/2)_a$&$1^{+-}$&2.68&4.03&5.15&6.14&7.01&7.80&8.53\\
$(1/2,1/2)_b$&$1^{--}$&2.78&4.18&5.32&6.30&7.17&7.96&8.68\\
$(0,1)\oplus(1,0)$&$1^{--}$&1.55&3.28&4.56&5.64&6.57&7.40&8.16\\
$(0,1)\oplus(1,0)$&$1^{++}$&2.20&3.73&4.95&5.98&6.88&7.69&8.43\\
\hline
$(1/2,1/2)_a$&$2^{-+}$&3.89&4.98&5.94&6.80&7.59&8.31&8.99\\
$(1/2,1/2)_b$&$2^{++}$&3.91&5.02&6.00&6.88&7.67&8.41&9.09\\
$(0,1)\oplus(1,0)$&$2^{++}$&3.60&4.67&5.64&6.51&7.31&8.06&8.75\\
$(0,1)\oplus(1,0)$&$2^{--}$&3.67&4.80&5.80&6.68&7.49&8.23&8.91\\
\hline
$(1/2,1/2)_a$&$3^{+-}$&4.82&5.77&6.62&7.41&8.13&8.81&9.45\\
$(1/2,1/2)_b$&$3^{--}$&4.82&5.78&6.65&7.44&8.17&8.86&9.50\\
$(0,1)\oplus(1,0)$&$3^{--}$&4.68&5.63&6.48&7.26&7.99&8.67&9.30\\
$(0,1)\oplus(1,0)$&$3^{++}$&4.69&5.66&6.53&7.32&8.06&8.75&9.39\\
\hline
$(1/2,1/2)_a$&$4^{-+}$&5.59&6.45&7.23&7.96&8.64&9.28&9.89\\
$(1/2,1/2)_b$&$4^{++}$&5.59&6.45&7.24&7.97&8.66&9.30&9.92\\
$(0,1)\oplus(1,0)$&$4^{++}$&5.51&6.36&7.15&7.88&8.56&9.20&9.80\\
$(0,1)\oplus(1,0)$&$4^{--}$&5.51&6.37&7.16&7.90&8.58&9.23&9.84\\
\hline
$(1/2,1/2)_a$&$5^{+-}$&6.27&7.05&7.78&8.47&9.11&9.72&10.3\\
$(1/2,1/2)_b$&$5^{--}$&6.27&7.06&7.79&8.47&9.12&9.73&10.3\\
$(0,1)\oplus(1,0)$&$5^{--}$&6.21&7.00&7.73&8.41&9.06&9.67&10.3\\
$(0,1)\oplus(1,0)$&$5^{++}$&6.21&7.00&7.73&8.42&9.07&9.68&10.3\\
\hline
$(1/2,1/2)_a$&$6^{-+}$&6.88&7.61&8.29&8.94&9.55&10.1&10.7\\
$(1/2,1/2)_b$&$6^{++}$&6.88&7.61&8.29&8.94&9.56&10.1&10.7\\
$(0,1)\oplus(1,0)$&$6^{++}$&6.83&7.57&8.25&8.90&9.51&10.1&10.7\\
$(0,1)\oplus(1,0)$&$6^{--}$&6.83&7.57&8.26&8.90&9.52&10.1&10.7\\
\hline\hline
\end{tabular}
\end{table}

 When the chiral symmetry breaking Lorentz-scalar dynamical
mass of quarks is zero, then there are independent Bethe-Salpeter
amplitudes just according to the chiral representations (10)-(12).
A finite dynamical mass plays a role of the off-diagonal
matrix element and mixes the otherwise independent chiral Bethe-Salpeter
amplitudes for the states $1^{--},2^{++},3^{--},...$.
A key feature of this dynamical mass is that it is strongly
momentum-dependent and vanishes very fast once the momentum is
increased. When one increases excitation energy of a hadron,
one also increases a typical momentum of valence quarks. Consequently,
the chiral symmetry violating dynamical mass of quarks becomes small.
Hence the mixing of the independent chiral Bethe-Salpeter amplitudes
becomes small. A given state in the table is then assigned to
the chiral representation according to the chiral Bethe-Salpeter
amplitude that dominates in the given state.

In Fig. 2 the rates of the symmetry restoration against the radial
quantum number $n$ and spin $J$ are shown. It is seen that with the fixed
$J$ the splitting within the multiplets $\Delta M$ decreases asymptotically as
$1/\sqrt n$, dictated by the asymptotic linearity of the radial Regge
trajectories. This property is consistent with the dominance of the free
quark loop logarithm at short distances.

\begin{figure}
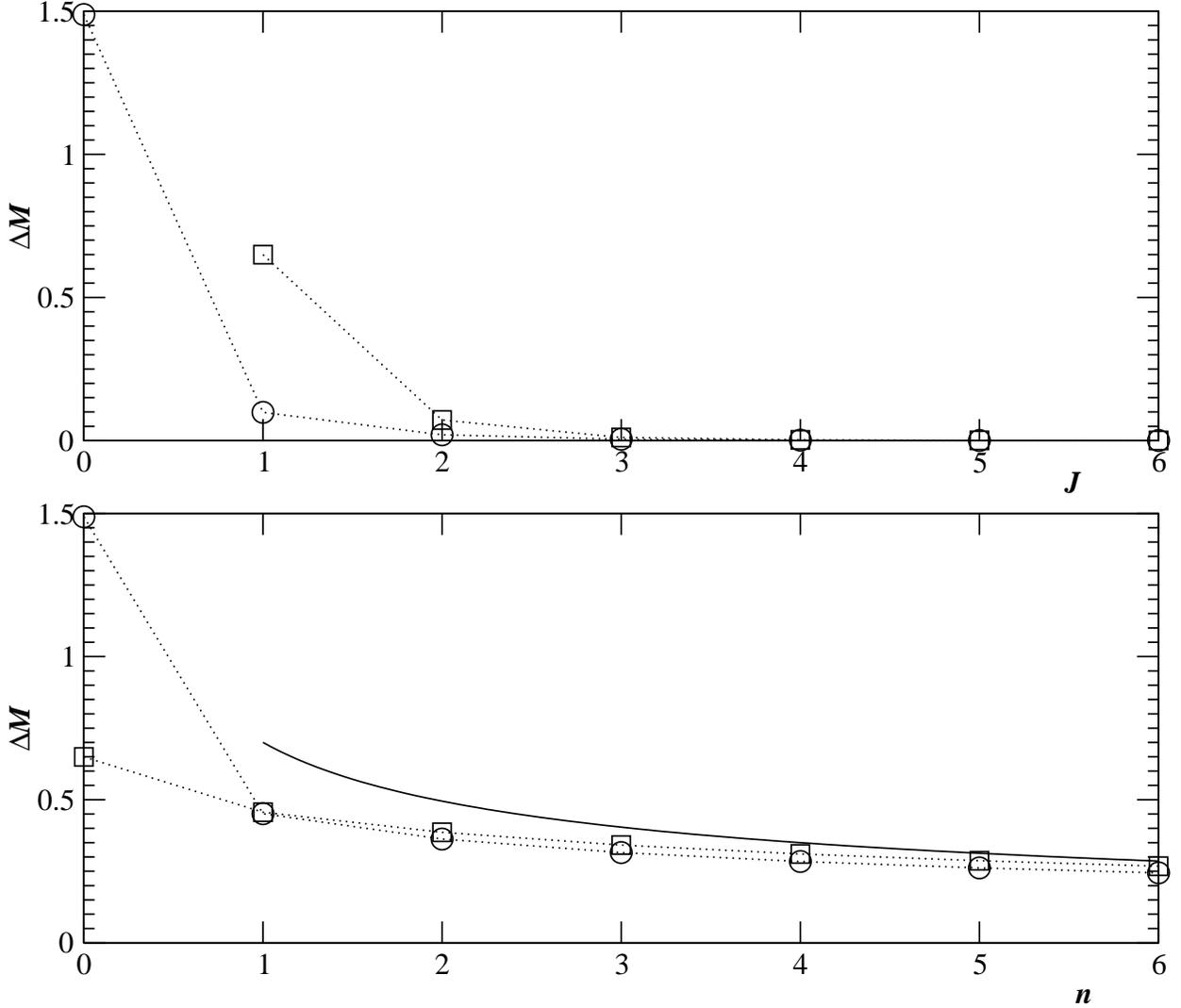

\includegraphics[width=\hsize,clip=]{delta_m_spin.eps}
\includegraphics[width=\hsize,clip=]{delta_m_radial.eps}
\caption{Mass splittings in units of $\sqrt{\sigma}$ for isovector mesons of the chiral
multiplets $(1/2,1/2)_a$ and $(1/2,1/2)_b$ (circles) and within
the multiplet $(0,1)\oplus(1,0)$ (squares) against $J$ for $n=0$ (top) and
against $n$ for $J=0$ and $J=1$, respectively (bottom).
The full line in the bottom plot is $0.7\sqrt{\sigma/n}$.}
\end{figure}

In Fig. 3 the angular and radial Regge trajectories are shown.
Both kinds of trajectories exhibit deviations from the linear behavior. 
This fact is obviously related to the
chiral symmetry breaking effects for lower mesons.
Note, that the chiral symmetry requires a doubling of some
of the radial and angular Regge trajectories for $J=1,2,..$. This is a
highly nontrivial prediction of chiral symmetry.
 For example, some of the
rho-mesons lie on the trajectory that is characterized by the chiral
index (0,1)+(1,0), while the other fit the trajectory with the
chiral index $(1/2,1/2)_b$. The intercepts of the asymptotic angular
Regge trajectories for mesons in the given and different 
chiral representations coincide.
Hence the asymptotic rate of the symmetry restoration with $J$ is {\it
faster} than $1/\sqrt J$.

\begin{figure}
\includegraphics[width=\hsize,clip=]{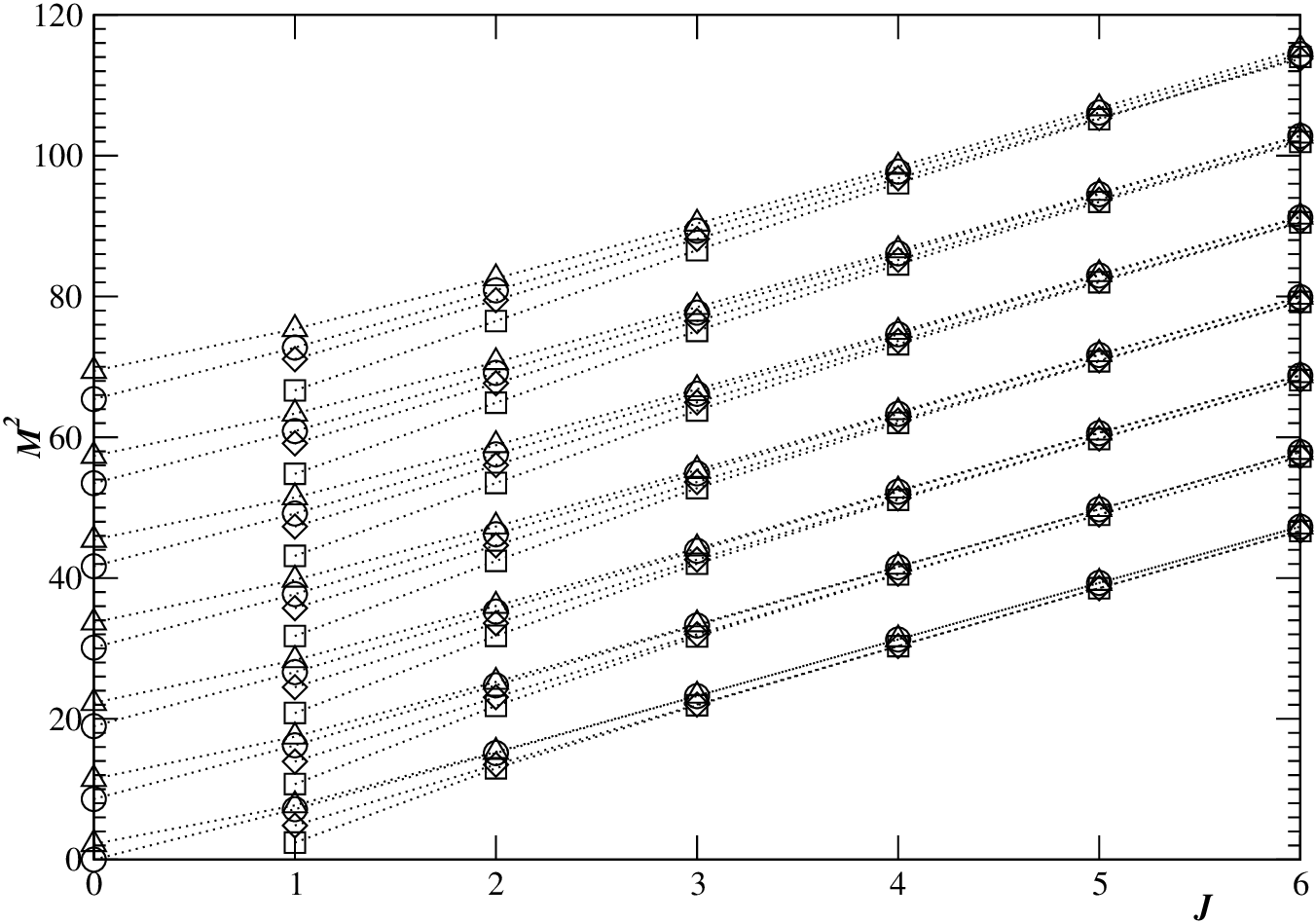}
\includegraphics[width=\hsize,clip=]{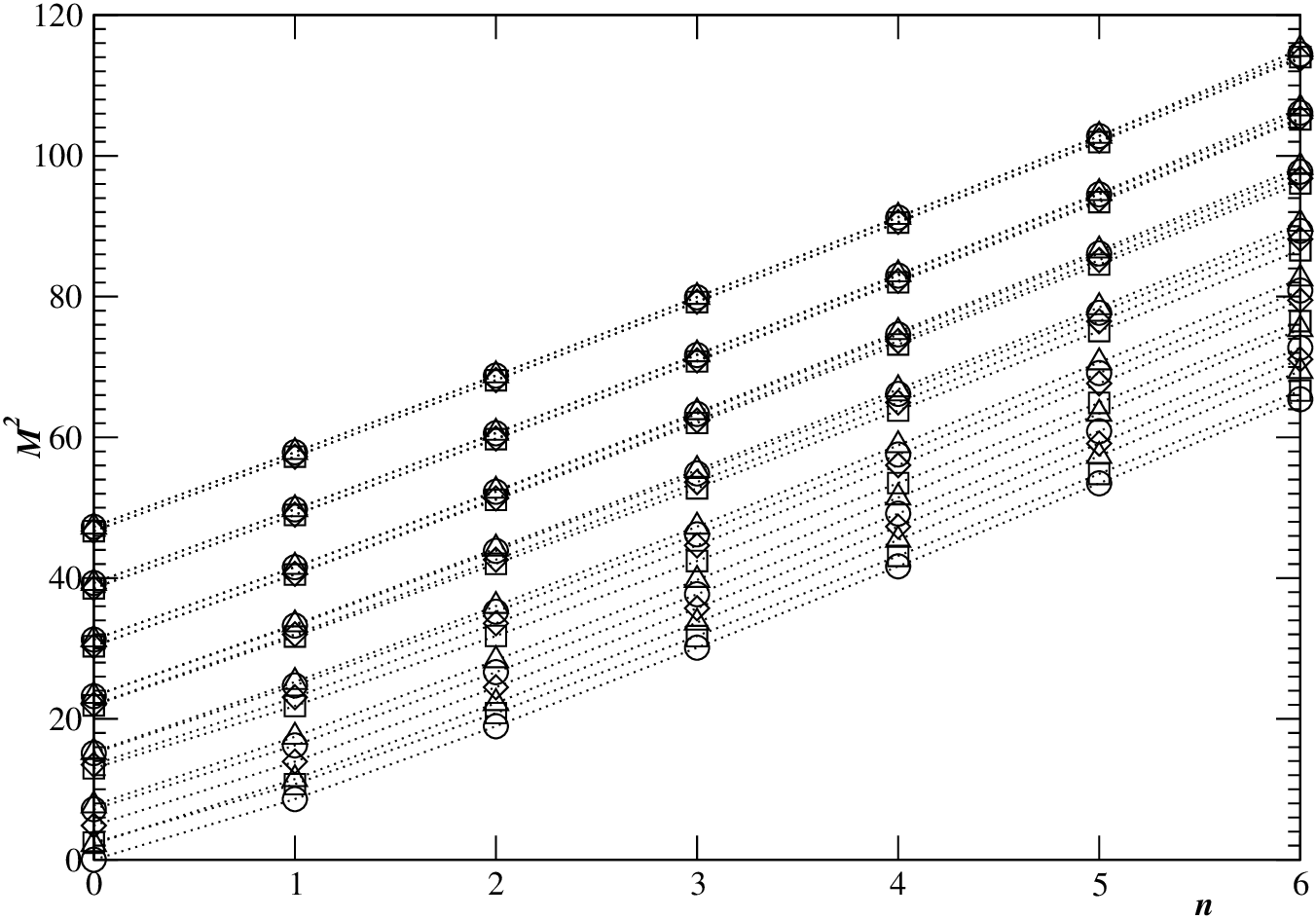}
\caption{Angular (top) and radial (bottom) Regge trajectories for isovector mesons
with $M^2$ in units of $\sigma$.
Mesons of the chiral multiplet $(1/2,1/2)_a$ are indicated by circles,
of $(1/2,1/2)_b$ by triangles, and of $(0,1)\oplus(1,0)$
by squares ($J^{++}$ and $J^{--}$ for even and odd $J$, respectively)
and diamonds ($J^{--}$ and $J^{++}$ for even and odd $J$, respectively).
}
\end{figure}

The  numerical result for the quark condensate is
$\langle\bar{q}q\rangle=(-0.231\sqrt{\sigma})^3$, which agrees
with the previous studies within the same model.
If we fix the string tension from the phenomenological
angular Regge trajectories, then $\sqrt{\sigma}\approx 300$ -- $400$ MeV and hence the quark
condensate is between $(-70\ \mbox{MeV})^3$ and $(-90\ \mbox{MeV})^3$ which
obviously underestimates the phenomenological value.
Probably this indicates
that other gluonic interactions could also contribute to chiral symmetry
breaking.
Notice that the string tension in Coulomb gauge can be larger than
the asymptotic one. Lattice results suggest a value about twice the value obtained here
\cite{Greensite:2004ke,Nakamura:2005ux}.
This would increase the value for the condensate but on the other hand lead to
an unrealistically small pion decay constant \cite{Adler:1984ri,Alkofer:1988tc}.

In  the limit $n \rightarrow \infty$ and/or  $J \rightarrow \infty$ 
one observes a complete degeneracy of all multiplets, which means
that the states fall into 
$$ [(0,1/2) \oplus (1/2,0)] \times [(0,1/2) \oplus (1/2,0)]$$
%
representation that combines all possible chiral representations
for the systems of two massless quarks \cite{G4}. This means that in this limit the
loop effects disappear completely and the system becomes classical \cite{G5,G6}.

\begin{acknowledgments}

RFW acknowledges helpful discussions with R. Alkofer, M. Kloker,
and A. Krassnigg. 
This work was supported by: Deutsche Forschungsgemeinschaft
(project Al 279/5-1) and the Austrian Science Fund (projects P16823-N08
and P19168-N16).
\end{acknowledgments}

\end{document}